\documentclass{jfm} 
 
\usepackage{amsmath} 
\usepackage{graphicx} 
\usepackage{color} 
\usepackage{upgreek} 
 
\usepackage{bm}

\title{Stirring by Periodic Arrays of Microswimmers} 
 
\author{Joost de Graaf\aff{1} 
  \corresp{\email{j.degraaf@ed.ac.uk}} 
 \and Joakim Stenhammar\aff{2}} 
\affiliation{\aff{1}SUPA, School of Physics and Astronomy, The University of Edinburgh, King's Buildings, Peter Guthrie Tait Road, Edinburgh, EH9 3FD, United Kingdom 
\aff{2}Division of Physical Chemistry, Lund University, P.O. Box 124, S-221 00 Lund, Sweden} 
 
\begin{document} 
 
\maketitle 
 
\begin{abstract} 
The interaction between swimming microorganisms or artificial self-propelled colloids and passive (tracer) particles in a fluid leads to enhanced diffusion of the tracers. This enhancement has attracted strong interest, as it could lead to new strategies to tackle the difficult problem of mixing on a microfluidic scale. Most of the theoretical work on this topic has focused on hydrodynamic interactions between the tracers and swimmers in a bulk fluid. However, in simulations, periodic boundary conditions (PBCs) are often imposed on the sample and the fluid. Here, we theoretically analyze the effect of PBCs on the hydrodynamic interactions between tracer particles and microswimmers. We formulate an Ewald sum for the leading-order stresslet singularity produced by a swimmer to probe the effect of PBCs on tracer trajectories. We find that introducing periodicity into the system has a surprisingly significant effect, even for relatively small swimmer-tracer separations. We also find that the bulk limit is only reached for very large system sizes, which are challenging to simulate with most hydrodynamic solvers. 
\end{abstract} 
 
\section{\label{sec:intro}Introduction} 
 
The physical properties of dense and dilute suspensions of biological as well as artificial ``microswimmers'' has attracted significant attention over the last few decades (\cite{Marchetti-2013}). Many of the experimental (\cite{Wu-2000,Leptos-2009,Valeriani-2011,Jepson-2013,Mino-2013,Jeanneret-2016}) and theoretical (\cite{Underhill-2008,Dunkel-2010,Thiffeault-2010,Pedley-2010,Lin-2011,Pushkin13mix,Pushkin-2013,Morozov-2014,Krishnamurthy-2015,Thiffeault-2015}) studies on biological microswimmer suspensions have focused on the enhanced diffusion of passive (\textit{i.e.}, non-swimming) tracer particles immersed in bacterial or algal suspensions. This strongly enhanced tracer diffusion, and especially its linear dependence on microswimmer density, has been rationalized in terms of the tracers being advected by a superposition of the hydrodynamic far-fields of the swimming microorganisms~(\cite{Pushkin13mix,Pushkin-2013}). Microorganisms are self-propelled rather than externally forced (provided one neglects the effects of gravity), and therefore the leading-order flow-field singularity of such an organism is that of a hydrodynamic dipole (stresslet)~(\cite{Drescher-2010,Drescher-2011}). In the simplest picture, mainly relevant to swimming bacteria such as \textit{E. coli}, the stresslet can be seen as being composed by one propulsive force acting on the fluid, and an equal and opposite force at the other end, coming from the drag imposed by the fluid on the swimmer. Depending on the details of the propulsion mechanism, the stresslet can have either symmetry: rear-actuated organisms such as~\textit{E. coli} and~\textit{Salmonella} bacteria are usually referred to as ``pushers'' (extensile), while front-actuated microswimmers such as the alga~\textit{Chlamydomonas} are referred to as ``pullers'' (contractile).  
 
When numerically simulating bulk suspensions and fluids, it is usually necessary to use periodic boundary conditions (PBCs) in order to emulate an infinite system. While obviating the need of solid confinement, the introduction of PBCs instead imposes an artificial periodicity on the system. This periodicity is especially important due to the slow $r^{-2}$ decay of the stresslet flow-field, identical to that of the electrostatic potential outside an electric dipole. For electrostatic systems, the artificial periodicity introduced by the use of PBCs has been shown to lead to spurious shifts of the thermodynamic and structural properties of both Coulombic (\cite{Hunenberger-1999}) and dipolar (\cite{Stenhammar-2011}) fluids. The slow decay of the stresslet flow field furthermore requires elaborate methods in order to analyze the resulting periodic sums, inspired by the original Ewald summation technique for charged systems (\cite{ewald1921berechnung,Wells-2015}). In this contribution, we will therefore analyze the effect of periodic boundary conditions (PBCs) on the advection of tracer particles caused by the flow field of microswimmers.  
 
\begin{figure} 
\centering  
  \includegraphics[scale=0.75]{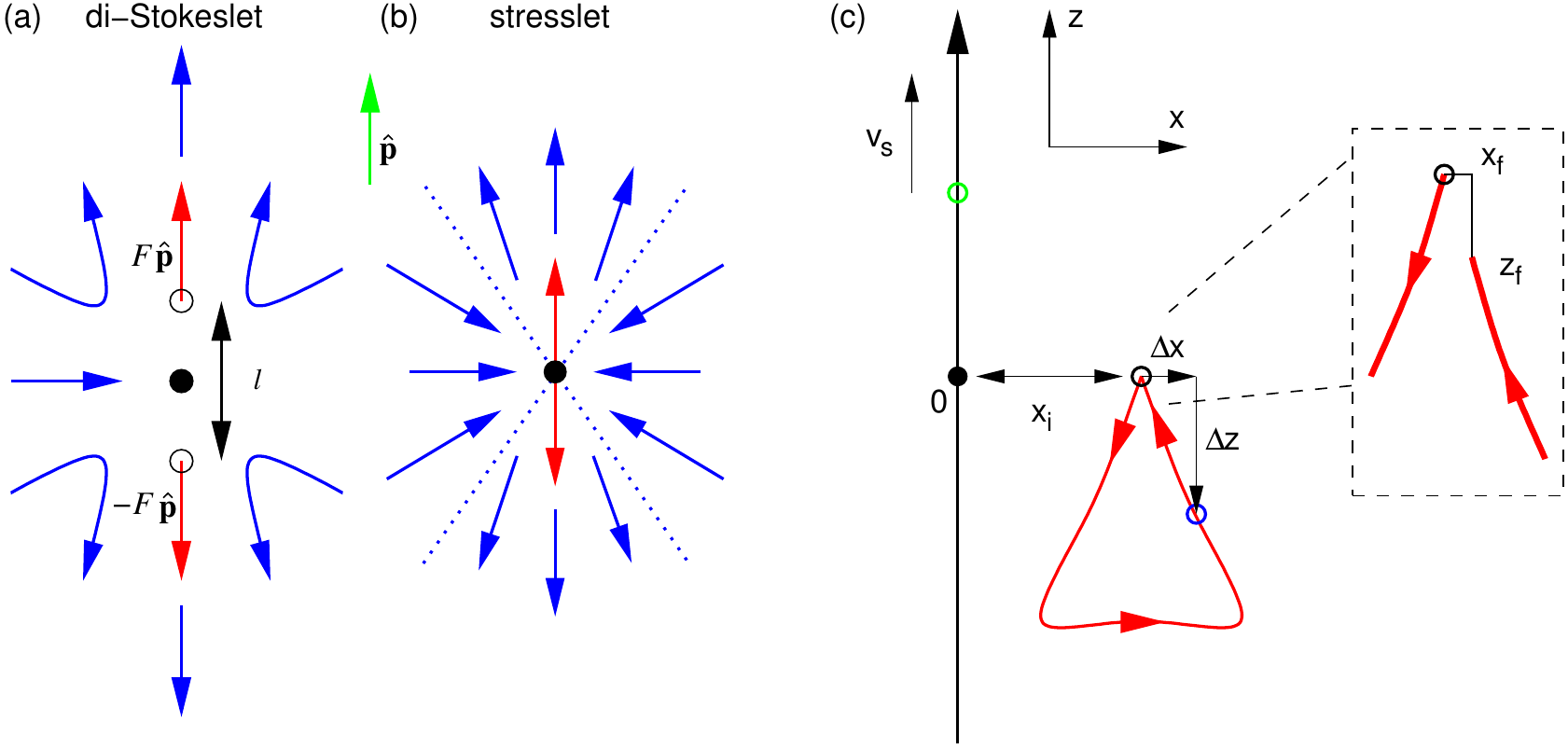}  
  \caption{\label{fig:distokes}(color online) The relevant geometric parameters. Sketch of the flow field around (a) a force-free extended pusher swimmer, the so-called ``di-Stokeslet'', and (b) the corresponding pusher point stresslet. The black dot indicates the center of the particle, the two open circles the positions where the force $F \boldsymbol{\hat{p}}$ and counter force $-F \boldsymbol{\hat{p}}$ are applied to the fluid (red arrows). $l$ denotes the dipole length, the green arrow shows the orientation $\boldsymbol{\hat{p}}$ of both swimmers, and the blue arrows give an impression of the stream lines around the swimmer. For the stresslet, the stream lines originate and terminate in the origin and are straight. The transition from outward to inward flow happens at a fixed angle ($\arccos (1/\sqrt{3})$), as indicated by the blue dotted lines. (c) Two-dimensional (2D) representation of the bulk geometry. The trajectory of the swimmer (green circle) is along the $z$-axis from $-\infty$ to $\infty$, which the swimmer traverses with speed $v_{s}$. The tracer (blue circle) is advected along the red trajectory -- here we have assumed a puller type swimmer. The initial position of the tracer (black circle) is given by $x_{i}\boldsymbol{\hat{x}}$. The tracer displacement due to advection by the swimmer flow field is measured using $\Delta x$ and $\Delta z$ which are measured with respect to $x_{i}$ and $z_{i}$. The final tracer position, as assumed once the swimmer has completed its trajectory, is given by $x_{f}$ and $z_{f}$, see the inset.}    
\end{figure}  
 
We start by considering the effect of swimmer-tracer separation and the difference between the extended ``di-Stokeslet'' and point stresslet descriptions of a microswimmer (see Fig.~\ref{fig:distokes}a,b) in an infinite (non-periodic) system. Here, we confirm that for small swimmer-tracer separations the tracers exhibit non-closed trajectories, leading to deviations from the standard concave triangular trajectories followed by tracers advected in dipolar flow fields (\cite{Pushkin-2013,Pushkin13mix}). Furthermore, in this regime, we find that the difference between the stresslet and di-Stokeslet descriptions becomes appreciable. Next, we extend the Ewald-summation method for Stokeslets put forward by \cite{Beenakker86} and detailed in \cite{Pozrikidis96} to properly include the effect of PBCs. We show that these significantly impact the trajectories of the tracers, even for small swimmer-tracer separation and relatively large boxes. Finally, we investigate the limit of large simulation boxes and find that unpractically large sizes -- from the perspective of simulations -- are needed for convergence to the bulk system behavior. Our work therefore raises questions concerning the suitability of common fluid-dynamic solvers to study the bulk behavior of microswimmers. 
 
\section{\label{sec:theory}Theory} 
 
The velocity field at a point $\boldsymbol{r}$ due to an extended stresslet (``di-Stokeslet'', see Fig.~\ref{fig:distokes}a) in bulk is given by 
\begin{align} 
\label{eq:Stressex} \boldsymbol{u}(\boldsymbol{r}) &= \frac{F}{8 \pi \mu} \left[ \mathsfbi{S} \Big( \boldsymbol{r} - \frac{l}{2} \boldsymbol{\hat{p}} \Big) - \mathsfbi{S} \Big( \boldsymbol{r} + \frac{l}{2} \boldsymbol{\hat{p}} \Big) \right] \boldsymbol{\hat{p}}, 
\end{align}  
where $\mu$ is the dynamic viscosity, $F$ the force magnitude, $l$ the point separation, $\boldsymbol{\hat{p}}$ the unit vector indicating the swimmer's orientation, and  
\begin{equation} 
\label{eq:Stokeslet} 
\mathsfbi{S}(\boldsymbol{r}) = \frac{1}{r} \left( \mathsfbi{I} + \boldsymbol{\hat{r}} \otimes \boldsymbol{\hat{r}} \right) 
\end{equation} 
the bulk Stokeslet, with $\mathsfbi{I}$ the identity matrix, $r \equiv \vert \boldsymbol{r} \vert$, $\boldsymbol{\hat{r}} \equiv \boldsymbol{r}/r$, and $\otimes$ the dyadic product. The corresponding expression for the (point) stresslet is obtained by taking the limit $l \rightarrow 0$, see Fig.~\ref{fig:distokes}b, yielding a directional derivative and the expression 
\begin{align}  
\label{eq:Stressudef1} \boldsymbol{u}(\boldsymbol{r}) &= -\frac{\kappa}{8 \pi \mu} \left[ \left( \boldsymbol{\hat{p}} \cdot \boldsymbol{\nabla} \right) \mathsfbi{S}(\boldsymbol{r}) \right] \boldsymbol{\hat{p}} = \frac{\kappa}{8\pi \mu r^2}(3(\boldsymbol{\hat{p}}\cdot \boldsymbol{\hat{r}})^2 - 1)\boldsymbol{\hat{r}}, 
\end{align}   
with $\boldsymbol{\nabla}$ the gradient with respect to $\boldsymbol{r}$. In our construction, positive values of the stresslet strength $\kappa = Fl$ correspond to pusher swimmers and negative values to puller swimmers. 
 
In a system with PBCs, the velocity experienced at $\boldsymbol{r}_{\boldsymbol{0}}$ in the central box due to the array of point forces $F \boldsymbol{\hat{p}}$ is formally given by 
\begin{align} 
\label{eq:Pernaive} \boldsymbol{u}(\boldsymbol{r}_{\boldsymbol{0}}) &= \frac{F}{8\pi \mu} \left( \sum_{\boldsymbol{n}} \mathsfbi{S}(\boldsymbol{r}_{\boldsymbol{n}}) \right) \boldsymbol{\hat{p}}, 
\end{align} 
where $\boldsymbol{r}_{\boldsymbol{n}} = \boldsymbol{r}_{\boldsymbol{0}}  - \boldsymbol{X}_{\boldsymbol{n}}$ and $\boldsymbol{X}_{\boldsymbol{n}} = n_{0} \boldsymbol{a} + n_{1} \boldsymbol{b} + n_{2} \boldsymbol{c}$, with $\boldsymbol{n} \in \mathbb{Z}^{3}$ and $\boldsymbol{a}$, $\boldsymbol{b}$, and $\boldsymbol{c}$ the sides of the box. However, the slow $1/r$ decay of $\mathsfbi{S}$ leads to the sum in Eq.~\eqref{eq:Pernaive} being conditionally convergent, just as when summing up electrostatic charge and dipole potentials. Thus, the flow field will depend on the order in which the terms are summed. From a physical perspective, however, one usually selects a summation over spherical (or, in the case of a non-cubic box, spheroidal) shells. From a computational viewpoint, the slow convergence of Eq.~\eqref{eq:Pernaive} furthermore makes it necessary to employ more elaborate methods, inspired by the Ewald summation used in electrostatics, to transform the sum into a more rapidly (and absolutely) convergent one.  
 
Here, we will use the formalism put forward by \cite{Beenakker86} and \cite{Pozrikidis96}, who suggested splitting the Stokeslet according to 
\begin{align} 
\label{eq:Stokesletnew} \mathsfbi{S}(\boldsymbol{r}) &= \boldsymbol{\Theta}(\boldsymbol{r}) + \boldsymbol{\Psi}(\boldsymbol{r}) , 
\end{align} 
with 
\begin{align} 
\label{eq:Stokesletreal} \boldsymbol{\Theta}(\boldsymbol{r}) &= 8\pi \left( \mathsfbi{I}\nabla^{2} - \boldsymbol{\nabla} \otimes \boldsymbol{\nabla} \right)B(r)\mathrm{erfc}(\xi r) ; \\ 
\label{eq:StokesletFour} \boldsymbol{\Psi}(\boldsymbol{r}) &= 8\pi \left( \mathsfbi{I}\nabla^{2} - \boldsymbol{\nabla} \otimes \boldsymbol{\nabla} \right)B(r)\mathrm{erf}(\xi r), 
\end{align} 
where $\nabla^{2}$ is the Laplacian, $B(r) = r/(8\pi)$ the fundamental solution of the biharmonic equation, $\mathrm{erf}(z)$ the error function, $\mathrm{erfc}(z) = 1 - \mathrm{erf}(z)$ its complement, and $\xi \geq 0$ is a splitting parameter. $\boldsymbol{\Theta}$ now yields a rapidly convergent sum in real space (dominant for $\xi \ll 1$), whereas the long-range part $\boldsymbol{\Psi}$ (dominant for $\xi \gg 1$) will be handled in Fourier space, where it will also become absolutely convergent. As shown by \cite{Pozrikidis96}, this splitting yields  
\begin{align} 
\label{eq:Theta} \boldsymbol{\Theta}(\boldsymbol{r}) &= \frac{1}{r}\left( C(\xi r)\mathsfbi{I} + D(\xi r)\boldsymbol{\hat{r}} \otimes \boldsymbol{\hat{r}} \right) ; \\ 
\label{eq:auxC} C(z) &= \mathrm{erfc}(z) + \frac{2z}{\sqrt{\pi}} \left( 2 z^{2} - 3 \right) e^{-z^{2}} ; \\ 
\label{eq:auxD} D(z) &= \mathrm{erfc}(z) + \frac{2z}{\sqrt{\pi}} \left( 1 - 2 z^{2} \right) e^{-z^{2}} . 
\end{align} 
and  
\begin{align} 
\label{eq:Psi} \check{\boldsymbol{\Psi}}(\boldsymbol{k}) &=  \frac{8\pi}{k^{2}} \left( \mathsfbi{I} - \boldsymbol{\hat{k}} \otimes \boldsymbol{\hat{k}} \right) \left( 1 + \frac{\omega^{2}}{4} + \frac{\omega^{4}}{8} \right) e^{-\omega^{2}/4} e^{-i \boldsymbol{k} \cdot \boldsymbol{r}_{\boldsymbol{0}}} , 
\end{align} 
where $\check{\boldsymbol{\Psi}}$ is the Fourier transform of $\boldsymbol{\Psi}$ (the ``check'' symbol is used throughout to denote the transform), $\boldsymbol{k}$ is the reciprocal-space wave vector, $k \equiv \vert \boldsymbol{k} \vert$, $\boldsymbol{\hat{k}} \equiv \boldsymbol{k}/k$, and $\omega \equiv k/\xi$. The sum in Eq.~\eqref{eq:Pernaive} is now given by the two absolutely convergent sums 
\begin{equation} 
\mathsfbi{S}_{\mathrm{PBC}}(\boldsymbol{r}) \equiv \sum_{\boldsymbol{n}} \mathsfbi{S}(\boldsymbol{r}_{\boldsymbol{n}}) = \sum_{\boldsymbol{n}} \boldsymbol{\Theta}(\boldsymbol{r}_{\boldsymbol{n}}) + \frac{1}{V} \sum_{\boldsymbol{m} \ne \boldsymbol{0}} \check{\boldsymbol{\Psi}}(\boldsymbol{k}_{\boldsymbol{m}}), 
\end{equation} 
where $V \equiv \boldsymbol{c} \cdot ( \boldsymbol{a} \times \boldsymbol{b} )$ is the system volume, $\boldsymbol{k}_{\boldsymbol{m}} \equiv 2 \pi m_{0} (\boldsymbol{b} \times \boldsymbol{c} )/V + 2 \pi m_{1} (\boldsymbol{c} \times \boldsymbol{a} )/V + 2 \pi m_{2} (\boldsymbol{a} \times \boldsymbol{b} )/V$ with $\boldsymbol{m} \in \mathbb{Z}^{3}$, and it is assumed that the left-hand side sum is carried out over spherical shells. Note also that the $\boldsymbol{k} = \boldsymbol{0}$ term should be excluded from the right-hand side to eliminate the net force as discussed in~\cite{Pozrikidis96}.  
 
The velocity field due to a periodic array of stresslets can now be obtained using a similar directional derivative as 
\begin{align} 
\label{eq:UPBCStress} \boldsymbol{u}_{\mathrm{PBC}}(\boldsymbol{r}) &= -\frac{\kappa}{8\pi\mu} [\left( \boldsymbol{\hat{p}} \cdot \boldsymbol{\nabla} \right) \mathsfbi{S}_{\mathrm{PBC}}(\boldsymbol{r})]  \boldsymbol{\hat{p}}. 
\end{align}  
The real-space part yields 
\begin{equation}\label{eq:Stresscalc} 
\left( \boldsymbol{\hat{p}} \cdot \boldsymbol{\nabla} \right) \boldsymbol{\Theta}(\boldsymbol{r}) \boldsymbol{\hat{p}} = \frac{1}{r^{2}} \Big[ D(\xi r) \boldsymbol{\hat{r}} - E(\xi r) (\boldsymbol{\hat{r}} \cdot \boldsymbol{\hat{p}})^{2} \boldsymbol{\hat{r}} - F(\xi r) (\boldsymbol{\hat{r}} \cdot \boldsymbol{\hat{p}}) \boldsymbol{\hat{p}}  \Big], 
\end{equation} 
where $C(z)$ and $D(z)$ are given by Eqs.~\eqref{eq:auxC} and~\eqref{eq:auxD}, and we have defined the new auxiliary functions 
\begin{align} 
\label{eq:auxE} E(z) &\equiv 3D(z) - zD'(z) = 3 \mathrm{erfc}(z) + \frac{2z}{\sqrt{\pi}} \left( 3 + 2 z^{2} - 4 z^{4} \right) e^{-z^{2}}; \\ 
\label{eq:auxF} F(z) &\equiv C(z) - z C'(z) - D(z) = \frac{2z}{\sqrt{\pi}} \left( 4 z^{4} - 8 z^{2} \right) e^{-z^{2}} . 
\end{align} 
The Fourier-space part is similarly obtained from Eq.~\eqref{eq:Psi} and reads 
\begin{align} 
\label{eq:Psistress} \left( \boldsymbol{\hat{p}} \cdot \boldsymbol{\nabla} \right) \check{\boldsymbol{\Psi}}(\boldsymbol{k}) \boldsymbol{\hat{p}} &= -i \left( \boldsymbol{\hat{p}} \cdot \boldsymbol{\hat{k}} \right) k \check{\boldsymbol{\Psi}}(\boldsymbol{k}) \boldsymbol{\hat{p}} . 
\end{align} 
Thus, we may write for the velocity field due to a periodic array of stresslets with dipole strength $\kappa$ and direction $\boldsymbol{\hat{p}}$ 
\begin{align} 
\label{eq:uPBCstr} \boldsymbol{u}_{\mathrm{PBC}}(\boldsymbol{r}) &=  \sum_{\boldsymbol{n}} \boldsymbol{u}_{\mathrm{R}}(\boldsymbol{r}_{\boldsymbol{n}}) + \frac{1}{V} \sum_{\boldsymbol{m} \ne \boldsymbol{0}} \boldsymbol{u}_{\mathrm{F}}(\boldsymbol{k}_{\boldsymbol{m}}) , 
\end{align} 
with 
\begin{equation}\label{eq:ureal} 
\boldsymbol{u}_{\mathrm{R}}(\boldsymbol{r}) \equiv \frac{\kappa}{8\pi\mu r^{2}}\Big[E(\xi r) (\boldsymbol{\hat{r}} \cdot \boldsymbol{\hat{p}})^{2}  \boldsymbol{\hat{r}} + F(\xi r) (\boldsymbol{\hat{r}} \cdot \boldsymbol{\hat{p}}) \boldsymbol{\hat{p}}-D(\xi r) \boldsymbol{\hat{r}} \Big], 
\end{equation} 
and  
\begin{equation} 
\label{eq:ufour} \boldsymbol{u}_{\mathrm{F}}(\boldsymbol{k}) \equiv \frac{i\kappa}{\mu k} \left( \boldsymbol{\hat{p}} \cdot \boldsymbol{\hat{k}} \right) \left( \mathsfbi{I} - \boldsymbol{\hat{k}} \otimes \boldsymbol{\hat{k}} \right) \left( 1 + \frac{\omega^{2}}{4} + \frac{\omega^{4}}{8} \right) e^{-\omega^{2}/4} e^{-i \boldsymbol{k} \cdot \boldsymbol{r}_{\boldsymbol{0}}} \boldsymbol{\hat{p}}, 
\end{equation} 
with $D$, $E$, and $F$ as defined before. We finally note that these expressions are different from those derived in \cite{Klinteberg-2014}, as those results correspond to a periodic array of singularities that exclude the source term (the second term inside the bracket of Eq.~\eqref{eq:Stressudef1}).  
 
\section{\label{sec:approx}Comparison between di-Stokeslet and Stresslet Descriptions} 
 
Before assessing the effect of periodic boundaries on tracer trajectories, we will investigate the difference between the extended di-Stokeslet and point stresslet descriptions in the near-field of an infinite (non-periodic) system. To this end, we follow the strategy of \cite{Pushkin-2013} and perform a numerical calculation of the effect of a swimmer passing by a tracer particle, with the swimmer moving from $-\infty$ to $+\infty$ along the $z$-axis with constant velocity $v_{s}$, see Fig.~\ref{fig:distokes}c. Over the course of its trajectory, the swimmer interacts with a tracer initially positioned at $\boldsymbol{r}_{i} = (x_{i},0,0)$. Note that, by symmetry, the tracer motion may be restricted to the $xz$-plane. The tracer trajectory is parametrized by $\Delta x = x(t) - x_{i}$ and $\Delta z = z(t) - z_{i} = z(t)$, and is simply given by the advection it experiences due to the swimmer flow field. These assumptions result in the differential equations 
\begin{align} 
\label{eq:tracer}  \dot{\boldsymbol{r}}_{t}(t) &= \boldsymbol{u}(\boldsymbol{r}_{t}(t) - \boldsymbol{r}_{s}(t)); \\ 
\label{eq:swimmer} \dot{\boldsymbol{r}}_{s}(t) &= v_{s} \boldsymbol{\hat{z}}, 
\end{align} 
where $\boldsymbol{r}_{t}$ and $\boldsymbol{r}_{s}$ denote the tracer and swimmer positions, respectively. The fluid velocity $\boldsymbol{u}(\boldsymbol{r})$ in bulk is given either by the stresslet expression of Eq.~\eqref{eq:Stressudef1}, or its di-Stokeslet equivalent, Eq.~\eqref{eq:Stressex}. In the numerics, the swimmer trajectories start at $z = -5.0 \cdot 10^{4}$ and end at $z =  5.0 \cdot 10^{4}$, using a varying time-step size to ensure an accurate integration when the separation between the swimmer and tracer is small. We further assume $\mu = 1.0$ in our dimension-free units, and the swim-speed is kept constant at $v_{s} = 1.0\cdot10^{-3}$. We set $\vert \kappa \vert = 1.0$ in the stresslet model and $l = F = 1.0$ in the di-Stokeslet model unless otherwise specified. Finally, we denote the final position of the tracer by $\boldsymbol{r}_{f} = (x_{f},0,z_{f})$. 
 
\begin{figure*} 
\centering  
  \includegraphics[scale=0.75]{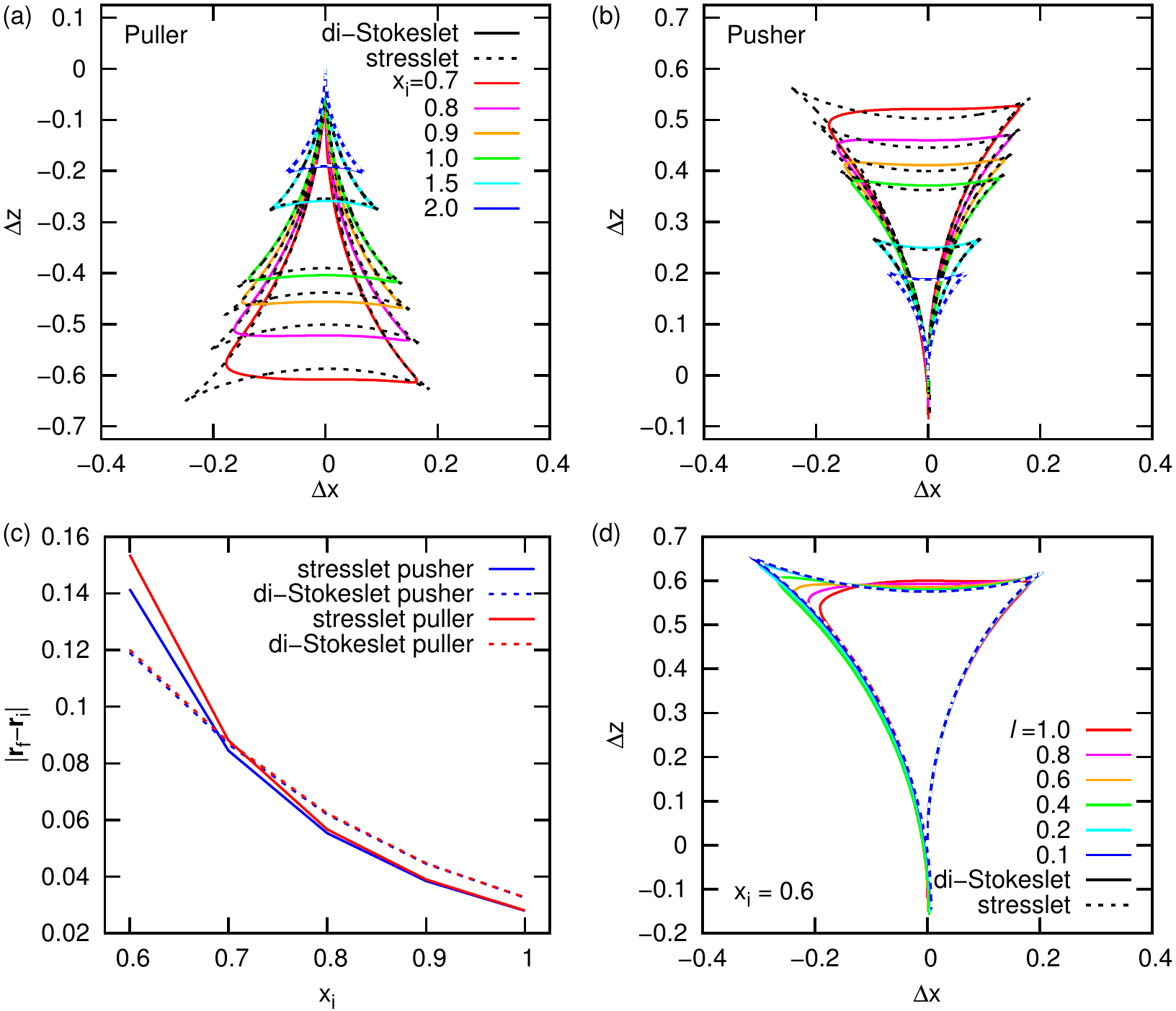}  
  \caption{\label{fig:small}(color online) Near-field trajectories of tracers advected by (a) pusher and (b) puller swimmers in an infinite (bulk) system. The results for the di-Stokeslet ($l = 1$) are given by the solid curves and for the stresslet by the dashed curves. (c) The net tracer displacement by a stresslet swimmer. The difference between the initial and final positions of the tracer $\vert \boldsymbol{r}_{f} - \boldsymbol{r}_{i} \vert$ after the swimmer has completed its trajectory is given as a function of the initial tracer position $x_{i}$. The results for pushers are shown in blue and for pullers in red. For both swimmer types the sign of the displacement is the same: negative for the $z$ coordinate and positive for the $x$ coordinate. (d) Comparison between the tracer trajectories caused by a pusher di-Stokeslet (solid curves) and stresslet (dashed curves) in bulk as a function of the di-Stokeslet length $l$ (keeping $\kappa = 1.0$ constant) for $x_{i} = 0.6$.}    
\end{figure*}

Figure~\ref{fig:small}(a,b) shows the difference between the di-Stokeslet and stresslet descriptions on the advection of tracer particles. For $x_{i} > l$ the deviation between the trajectories is minimal. However, for smaller values of the initial separation there is a clear ``rounding'' of the di-Stokeslet result with respect to that of the stresslet, which is a manifestation of the finite separation between the two force points. Clearly, the effect of the swimmer on the tracer becomes more pronounced as their separation is reduced. Below a certain value of the separation ($x_{i} \lesssim 0.5$ for our parameters) the tracer will start interacting with the ``naked'' singularity at the center of the swimmer -- or the front singularity in the di-Stokeslet approximation -- leading to numerical divergences in the tracer trajectories. These clearly require regularization, and could be prevented using additional swimmer-tracer interactions accounting for near-field lubrication effects and non-hydrodynamic interaction potentials; this is not considered here.  
 
It is also clear from Fig.~\ref{fig:small}(a,b) that the trajectory due to the stresslet swimmer becomes increasingly asymmetric with decreasing $x_{i}$. This can be explained by the tracer being increasingly subjected to the near-field flow of the swimmer. Both in the near and far field, the Darwin drift (\cite{Darwin-1953,Pushkin-2013}) leads to the tracer trajectories not being closed, yielding finite values of $\Delta x$ and $\Delta z$ after the swimmer has moved along its full trajectory. The effect is illustrated in Fig.~\ref{fig:small}c, where the difference $\vert \boldsymbol{r}_f - \boldsymbol{r}_{i} \vert$ between the initial and final position of the tracer is given as a function of $x_{i}$. As has been pointed out by \cite{Pushkin-2013}, this difference will approach zero in the limit where the tracer advection is negligible compared to the swimmer movement,~\textit{i.e.}, $\vert \dot{\boldsymbol{r}}_{t} \vert / \vert \dot{\boldsymbol{r}}_{s} \vert \rightarrow 0$, which corresponds to the limit $x_{i} \rightarrow \infty$. However, for the short swimmer-tracer separations studied in Fig.~\ref{fig:small}, this displacement is significantly different from zero. Furthermore, we note that the difference in the net displacement between the stresslet and di-Stokeslet descriptions is surprisingly small, even though the overall near-field trajectories of these two objects differ significantly.  
 
Finally, we demonstrate that the di-Stokeslet results reduce to those of the point stresslet in the limit $l \rightarrow 0$, as expected. Figure~\ref{fig:small}d shows this effect for $x_{i} = 0.6$ and various values of $l$ in the case of a pusher swimmer. For this small swimmer-tracer separation, a value of $l \approx 0.1$ has to be chosen for reasonable convergence to the stresslet result.  
 
\section{\label{sec:geom}The Effect of Periodicity on Tracer Trajectories} 
 
\begin{figure*} 
\centering  
  \includegraphics[scale=0.75]{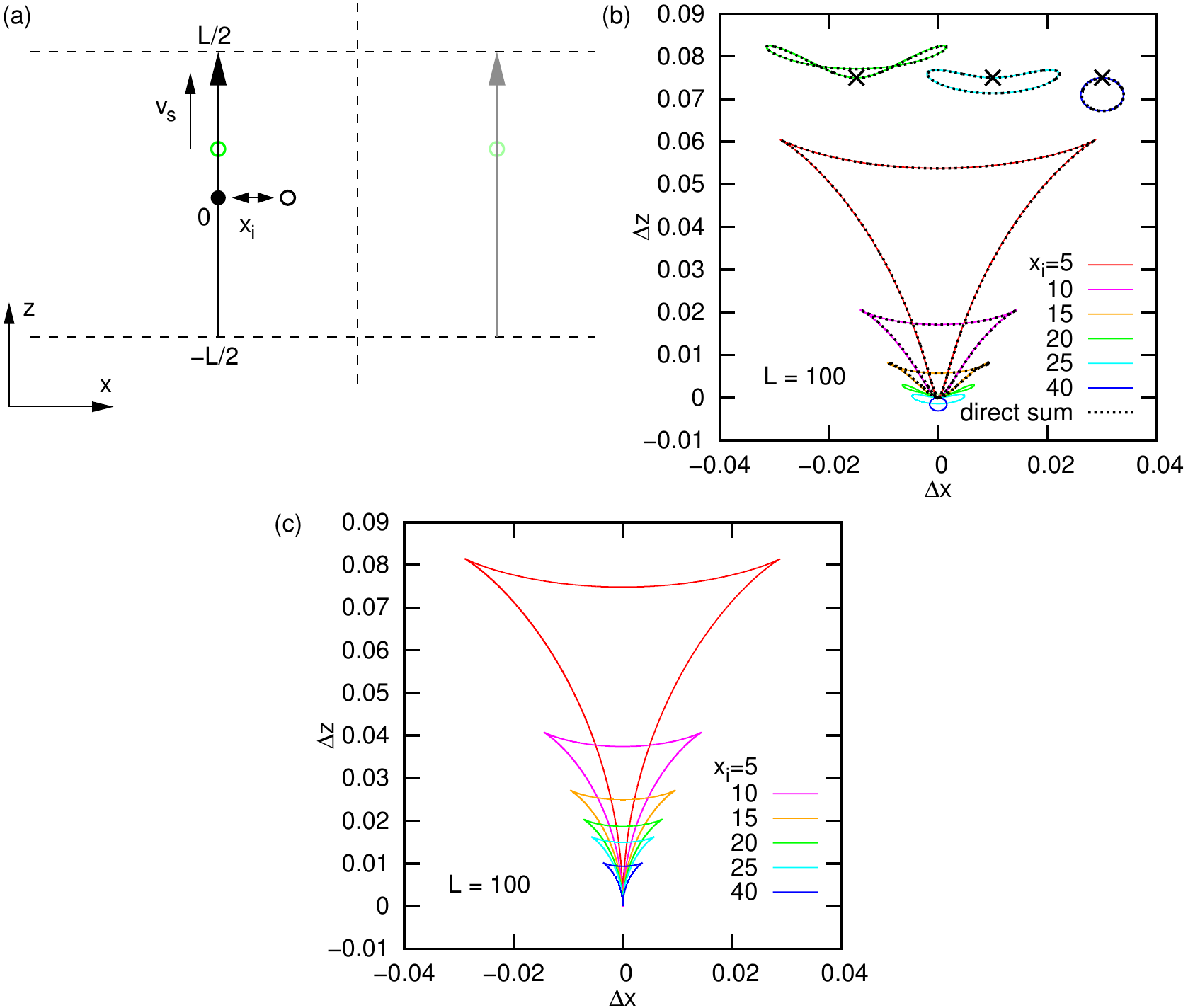}  
  \caption{\label{fig:pbcs}(color online) The effect of periodic boundary conditions on the advection of tracers by microswimmers. (a) 2D representation of a cubic simulation box with PBCs (dashed black lines). The swimmer (green circle) moves at speed $v_{s}$ along the $z$-axis from $-L/2$ to $L/2$, with $L$ the edge length of the box. The initial position $x_{i}\boldsymbol{\hat{x}}$ of the tracer is given by the black circle. One of the periodic duplicates of the system is shown using desaturated colors on the right-hand side. (b) Tracer trajectories in a cubic system with PBCs of edge length $L = 100$, for which the initial tracer-swimmer separation $x_{i}$ is varied. The three insets show the result for $x_{i} = 20$, $25$, and $40$ (from left to right) magnified by a factor of 2.5 and shifted with respect to the origin for clarity; the shifted origin is indicated using the $\times$ symbol. Solid lines show data obtained using the stresslet Ewald expressions, while dashed lines show the corresponding (partially converged) results from a direct summation in spherical shells. (c) Data corresponding to (b), but using the bulk stresslet expression. (d) The convergence towards the bulk tracer trajectory (dashed black curve) for a cubic system with PBCs (solid curves) for various values of the box length $L$. The initial separation is kept constant at $x_{i} = 10$.}    
\end{figure*}

We now assess the effect of periodic boundary conditions on the interaction between swimmers and tracers. The geometry in which we perform the calculation is sketched in Fig.~\ref{fig:pbcs}a. Our simulation box is cubic with edge length $L$, centered on the origin. The general numerical procedure is thus the same as that described in Section~\ref{sec:approx}, apart from that the swimmer now moves from $-L/2$ to $L/2$ along the $z$-axis. When integrating Eqs.~\eqref{eq:tracer}~and~\eqref{eq:swimmer}, the fluid flow velocity $\boldsymbol{u}(\boldsymbol{r})$ is now given by the PBC stresslet expression of Eq.~\eqref{eq:uPBCstr}. We furthermore used the value $\xi = \pi^{1/2}/L$ suggested by \cite{Beenakker86} for cubic boxes, and (cubic) real- and Fourier-space cutoffs of $\vert n_{\mathrm{max}} \vert$, $\vert m_{\mathrm{max}} \vert \leq 10$, which gave convergence to within the numerical rounding error with respect to a cutoff of $9$.  
 
Figure ~\ref{fig:pbcs}b shows tracer trajectories induced by a periodic array of pusher swimmers with identical properties to those discussed in Section~\ref{sec:approx}, for several values of $x_{i}$ in a box of length $L = 100$. We start by noticing that our Ewald expressions are in accordance with results of an explicit summation of stresslets in spherical shells (dashed lines in Fig.~\ref{fig:pbcs}b). It should be noted, however, that these sums are extremely slowly convergent, and that even for 40 layers of images (corresponding to $\approx$ 85 minutes of CPU time), the curves are not fully converged to the Ewald results, where the latter require only $\approx 10$ seconds of CPU time.  
 
Furthermore, it is clear that even for relatively small swimmer-tracer separations ($x_{i} = 5$), the observed trajectories are significantly perturbed by the use of PBCs. For long separations ($x_{i} \geq 20$), the trajectories are qualitatively wrong compared to the bulk results (Fig. \ref{fig:pbcs}c). Moreover, the PBC results underestimate the maximum value of $\Delta z$ by about 25\%, even for the shortest swimmer-tracer separations considered. While the overall PBC trajectories are still approximately closed (\textit{i.e.}, $\vert \boldsymbol{r}_{f} - \boldsymbol{r}_{i} \vert \approx 0$, when $r_{s} = L/2$), the observed discrepancy would have a large effect on tracer displacements for tumbling swimmers, where tracer trajectories are not closed.  
 
\begin{figure*} 
\centering  
  \includegraphics[scale=0.75]{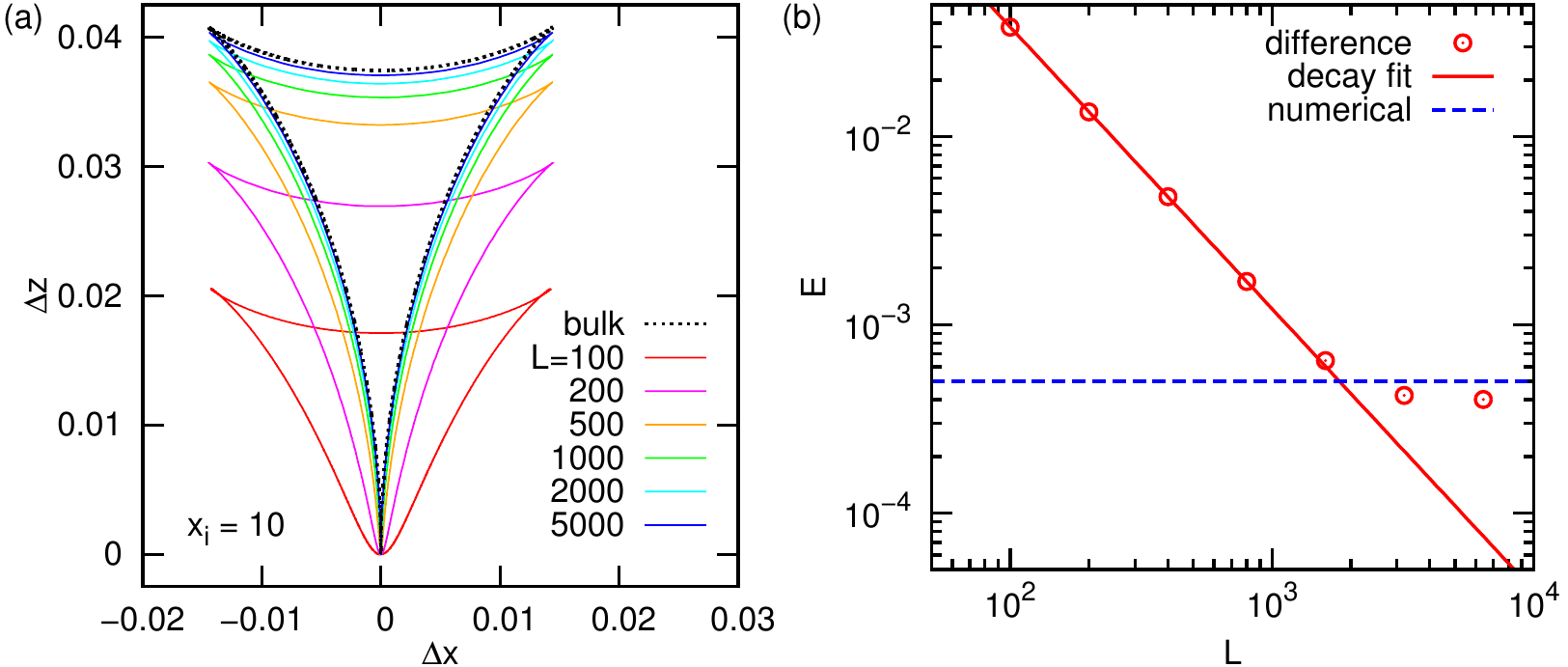}  
  \caption{\label{fig:fss}(color online) Finite size scaling for the swimmer-tracer advection and convergence of the flow field. (a) The convergence towards the bulk tracer trajectory (dashed black curve) for a cubic system with PBCs (solid curves) for various values of the box length $L$. The initial separation is kept constant at $x_{i} = 10$. (b) The near-field difference $E$ (red dots) between the bulk and PBCs flow field as a function of the box length, see Eq.~\ref{eq:err}. The solid red line is an extrapolation for the power-law decay in the error, while the blue dashed curve indicates the value of the error below which the difference becomes unreliable due to numerical rounding errors, as also evidenced by the kink in the graph.}    
\end{figure*} 
 
As a different manifestation of this effect, we consider the convergence towards the bulk behavior when increasing the box size while keeping $x_{i}$ constant at a value of 10 (Fig.~\ref{fig:fss}a). We note that, to obtain a reasonable correspondence with the bulk curve ($\approx 10\%$ error), a box length of $L = 1000$ is needed, corresponding to $L/x_{i} = 100$. This is a rather surprising and potentially worrying result, as this is far beyond the system sizes accessible in typical computer simulations. It is furthermore a much more long-ranged effect than what is observed in corresponding simulations of electric dipoles under PBCs, where the periodicity artifacts usually become negligible at length scales $\leq L/2$ (\cite{Stenhammar-2011}).  
 
To put this result in perspective, we should note that the near-field flow field around the swimmer converges more rapidly, as shown in Fig.~\ref{fig:fss}b, which shows the convergence of the fluid flow in PBCs onto that of the bulk stresslet. We computed the relative error $E$ in this flow-field reproduction as follows: a stresslet with our standard parameters was placed in the center of a cubic box with PBCs (or equivalently in bulk). We denote the stresslet-induced flow field in the cubic system with PBCs by $\boldsymbol{u}_{L}$ and the bulk flow field by $\boldsymbol{u}_{\infty}$. The error $E$ is then given by 
\begin{align} 
\label{eq:err} E &= \sqrt{ \frac{  \int \vert \boldsymbol{u}_{L}^{2} - \boldsymbol{u}_{\infty}^{2} \vert \mathrm{d}\boldsymbol{r} }{ \int \boldsymbol{u}_{\infty}^{2} \mathrm{d}\boldsymbol{r} } }, 
\end{align} 
where we integrate over the region $2 \le r \le 10$ to capture the region close to the swimmer. This definition has the advantage of being a relative error, weighted with the expected bulk local speed. That is, it gives an averaged ``per point'' error for the annular domain on which $E$ is evaluated. Convergence to $E = 0.1$ ($\approx 10\%$ error) is reached for $L < 100$, according to our definition; a different definition would yield more or less rapid convergence.  
 
By comparison our tracer trajectories are particularly sensitive to periodicity, since the swimmer traverses the length of the box, thereby always exposing the tracer to the strongly perturbed far field. However, if the physics of the system is partially governed by long-ranged interactions, as is the case for this specific example, then such slow convergence must be accounted for.  
 
\section{\label{sec:concl}Conclusions} 
 
The results shown here strongly highlight the effect of using periodic boundary conditions (PBCs) in hydrodynamic simulations (such as Stokesian dynamics (\cite{Lauga-2011}), dissipative particle dynamics (\cite{lugli11}), and lattice Boltzmann simulations (\cite{nash08,degraaf2016lattice})) of microswimmers. Specifically, they show that even on relatively small length-scales, the effect of using PBCs rather than a bulk solvent can be dramatic. Tracer advection for these swimmer-tracer separations is expected to contribute strongly to measureable properties such as the mean-square tracer displacement, which one should therefore expect to be significantly underestimated by the use of PBCs. Furthermore, the boundary effects highlighted here will also affect the interaction \textit{between} swimmers, thus having an impact on the modelling of collective behaviors in biofluids (\cite{Krishnamurthy-2015}). On the other hand, our results show that the difference between using extended or point stresslet dipoles is small for other than very short swimmer-tracer separations. In experiments, higher order singularities, thermal noise, lubrication effects, time-dependent motion (in the case of mechanical swimming), and non-hydrodynamic interactions are, however, likely to dominate at such small separations. In a broader context, this work highlights the importance of boundary and finite-size effects when studying systems with long-range fluid flows.  
 
\section*{\label{sec:acknowledgements}Acknowledgements} 
 
Helpful discussions with Mike Cates, Ludvig af Klinteberg, Alexander Morozov, and Cesare Nardini are kindly acknowledged. JdG thanks the ``Deutsche Forschungsgemeinschaft''  (DFG)  for funding  through  the  SPP 1726  ``Microswimmers: from  single  particle  motion  to  collective behavior'' and gratefully acknowledges funding by a Marie Sk{\l}odowska-Curie Intra European Fellowship (G.A. No. 654916) within Horizon 2020. JS is financed by a Project grant from the Swedish Research Council (2015-05449).  
 
\bibliographystyle{bib/jfm} 
\bibliography{bib/Manuscript} 
 
\end{document}